\newcommand{\aeff}{\Omega_\textrm{e}}

\newcommand{\ave}[1]{\langle #1 \rangle}
\newcommand{\degree}{\ensuremath{^\circ}}
\newcommand{\mkm}{\mathrm{km}}
\newcommand{\h}{\ensuremath{^\mathrm{h}}}
\newcommand{\m}{\ensuremath{^\mathrm{m}}}
\newcommand{\s}{\ensuremath{^\mathrm{s}}}

\documentclass{emulateapj}
\usepackage{graphicx}
\usepackage{natbib}
\usepackage{amssymb}

\begin{document}
\title{A Search for Occultations of Bright Stars by Small Kuiper Belt Objects using Megacam on the MMT}
\author{
F.~B.~Bianco\altaffilmark{1,2,3},
P.~Protopapas\altaffilmark{2,3},
B.~A.~McLeod\altaffilmark{2},
C.~R.~Alcock\altaffilmark{2},
M.~J.~Holman\altaffilmark{2},
M.~J.~Lehner\altaffilmark{4,1,2}
}
\altaffiltext{1}{University of Pennsylvania, 209 South 33rd Street, Philadelphia, PA 19104}
\email{fbianco@cfa.harvard.edu}
\altaffiltext{2}{Harvard-Smithsonian Center for Astrophysics, 60 Garden Street,
 Cambridge, MA 02138}
\altaffiltext{3}{Initiative in Innovative Computing at Harvard, 60 Oxford Street, Cambridge, MA 02138}
\altaffiltext{4}{Institute of Astronomy and Astrophysics, Academia Sinica.
 P.O. Box 23-141, Taipei 106, Taiwan}

\begin{abstract}
We conducted a search for occultations of bright stars by Kuiper Belt
Objects (KBOs) to estimate the density of sub-km KBOs in the sky.  We
report here the first results of this occultation survey of the
outer solar system conducted in June 2007 and
June/July 2008 at the MMT Observatory using Megacam, the large MMT
optical imager.  We used Megacam in a novel shutterless
\emph{continuous--readout} mode to achieve high precision photometry
at 200 Hz.  We present an analysis of $220$ star hours at
signal-to-noise ratio of 25 or greater. The survey efficiency is
greater than 10\% for occultations by KBOs of diameter
$d\ge0.7~\mathrm{km}$, and we report no detections in our dataset. We
set a new 95\% confidence level upper limit for the surface density
$\Sigma_N(d)$ of KBOs larger than 1 km: $\Sigma_N (d\ge 1~\mathrm{km})
~\leq~ 2.0 \times 10^8 ~\mathrm{deg}^{-2}$, and for KBOs larger than
0.7~km $\Sigma_N (d\ge 0.7~\mathrm{km}) ~ \leq ~ 4.8
\times 10^8 ~\mathrm{deg}^{-2}$.
\end{abstract}


\section{Introduction}
\setcounter{footnote}{0}
The size distribution of objects in the Kuiper Belt is believed to be
shaped by competitive processes of collisional agglomeration and
disruption.  The details of the structure of the Kuiper Belt size
distribution can reveal information on the internal structure of the
Kuiper Belt Objects (KBOs), the history of planet migration
\citep{2004AJ....128.1916K, 2005Icar..173..342P}, and the gas
history in the Solar System \citep{2009ApJ...690L.140K}. Large objects
in the Kuiper Belt (diameter $d \ge 30~\mathrm{km}$) can be observed
directly in reflected sunlight. The luminosity distribution for
objects larger than $~100~\mkm$ is well described by a single
power-law cumulative luminosity distribution $\Sigma_N(<R)=10^{\alpha
 (R-R_0)}$, where $\Sigma_N (<R)$ is the number of KBOs brighter than
magnitude R per degree in the sky on the ecliptic plane, with an index
$\alpha\sim0.7$ and $R_0\sim23$ \citep{2008arXiv0802.2285F,
 2008arXiv0804.3392F}.  This, under the assumption of 4\% constant
albedo, translates into a power-law size distribution $n(d) \propto
d^{-q}$ with power index $q \sim4.5$.  For these objects the size
distribution reflects the history of agglomeration.

There is strong evidence for a break in the slope of the distribution
at fainter magnitudes (smaller KBO sizes). Constraints on the
extrapolation of a single power law to magnitude greater than
$R\sim35$ were placed by \citet{2001ApJ...547L..69K}, who invoked
Olbers's Paradox applied to the Zodiacal Background, and by
\citet{2000AJ....119..945S}, who derived a slope for the distribution
of small KBOs of $q\approx3$ on the basis of the cratering on Triton as
observed by \emph{Voyager 2}. A further hint of a break in the size
distribution of KBOs is offered by the better--probed size
distribution of Jupiter Family Comets (JFCs, \citealt[and references
 therein]{2006Icar..182..527T}). These objects are likely injected
into their current orbits from the Kuiper Belt or the scattered disk
\citep{2008ApJ...687..714V}, and it is argued that their size
distribution would be preserved during this process. The size
distribution of JFCs is well represented by a shallow slope:
$q~=~2.7$ in the diameter range $1-10~\mkm$.  Future Microwave
Background surveys may also allow the setting of constraints on the
mass, distance, and size distribution of Outer Solar System (OSS)
objects \citep{2007ApJ...669.1406B}.

\citet{2004AJ....128.1364B} conducted a deep Hubble Space Telescope
survey with the Advanced Camera for Surveys which led to the detection
of three KBOs of magnitude $R > 26.5$; an extrapolation of the
bright end power-law would have predicted a factor of about $25$ more
detections for this survey. This reveals that a break in the power law
distribution must occur at magnitude brighter than $R~=~28.5$
($\mathrm{d}\sim20~\mkm$). This work remains the state of the art in
deep direct surveys of the OSS, with a completeness of
50\% at magnitude $R=28.5$. More recently \citet{2008arXiv0809.4166F}
and \citet{2009AJ....137...72F} detected more KBOs in the $R\le27$
region of the size spectrum and better constrained the slope of the
bright end of the power law and the location of the break, $R~\sim~25$
($d\sim50-100~\mkm$,  \citealt{2008arXiv0809.4166F},  \citealt{2009AJ....137...72F}).

\begin{figure}[ht!]
\centerline{\includegraphics[width=0.5\textwidth]{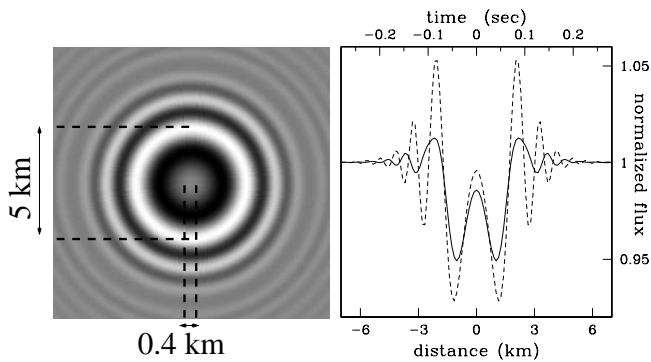}}
\caption[]{A simulated diffraction pattern (left panel) generated by a spherical $d~=~1~\mkm$ KBO occulting a magnitude 12 F0V star. The MMT/Megacam system bandpass (Sloan r' filter
and camera quantum efficiency) is assumed. The size of the KBO and the size of the Airy ring --~a measure of the cross section of the event~-- are shown for comparison. The right panel shows the diffraction signature of the event (assuming central crossing: impact parameter $b~=~0$) as a function of the distance to the point of closest approach (bottom scale). The top scale shows the time-line of the event assuming an observation conducted at opposition (relative velocity $v_\mathrm{rel}~\approx~25~\mathrm{km~s}^{-1}$). The occultation is sampled at 200~Hz (dashed line), and at 30~Hz, the \emph{effective} sampling rate after taking PSF effects into account (solid line, see Section~\ref{sec:noise}).}
\label{fig:occultationlcv}
\end{figure}

KBOs smaller than about $30~\mathrm{km}$ in diameter still elude
direct observations. Occultation surveys are the only observational
method presently expected to be able to detect such small objects in
the Kuiper Belt.  These surveys monitor background stars awaiting the
serendipitous alignment of KBOs with the stars.  The transit of a KBO
along the line of sight briefly modifies the observed flux of the
target star. At the distance of the Kuiper Belt ($\sim40~
\mathrm{AU}$) the size of the objects of interest is close to the
\emph{Fresnel scale} for visible light: this causes such occultation
events to be diffraction dominated phenomena. The Fresnel scale is
defined as $F = \sqrt{\lambda D/2}$ \citep{1980poet.book.....B,
  1987AJ.....93.1549R}, where $D$ is the distance to the occulter and
$\lambda$ the wavelength at which the occultation is observed.  In our
survey the bandpass of the observation is centered near $\lambda ~=
~500~\mathrm{nm}$ and, at distance $D~\approx~40~\mathrm{AU}$, $F
\approx 1.2~\mathrm{km}$.  Any occultation caused by objects in the
Kuiper Belt of a few kilometers in diameter or smaller will exhibit
prominent diffraction effects.  A diffraction pattern, characterized
by an alternation of bright and dark fringes centered on the KBO,
translates into a modulated lightcurve during the transit of the KBO
along the line of sight.  A unique feature, showing a series of
wiggles, and generally a reduction in flux, is imprinted in the time
series of the star \citep[see
  Figure~\ref{fig:occultationlcv}]{2000Icar..147..530R,
  2007AJ....134.1596N}.

The overall flux reduction is dominated by the size of the KBO, while
the duration of the event depends upon the relative velocity $v_\mathrm{rel}$ and the
size of the diffraction pattern $H$. We define $H$ as the diameter of
the first Airy ring, which it is limited by the Fresnel scale for sub-km
KBOs and by the size of the object for large KBOs as follows
\citep{2007AJ....134.1596N}:

\begin{equation}
H \approx \left[\left(2~\sqrt{3}F\right)^{3 \over 2}
+ d^{3 \over 2}\right]^{{2 \over 3}}~+~D\theta_\star,
\label{eq:width}
\end{equation}
where the additional $d\theta_\star$ term accounts for the finite
angular size of the star.  When observing at opposition the relative
velocity $v_\mathrm{rel}$ of an object orbiting the sun at
$40~\mathrm{AU}$ is about $25~\mathrm{km~s^{-1}}$ and the typical
duration of an occultation by sub-km KBOs is about $0.2$ seconds.

Occultation surveys were first proposed by
\citet{1976Natur.259..290B}, but only recently have results been
reported.  \cite{2007MNRAS.378.1287C} conducted a search for KBO
occultations in the archival RXTE X-ray observations of Scorpius-X1.
They reported a surprisingly high rate of occultation--like phenomena:
dips in the lightcurves compatible with occultations by objects
between $10$ and $200~\mathrm{m}$ in diameter.
\citet{2008ApJ...677.1241J} showed that most of the dips in the Sco-X1
lightcurves may be attributed to artificial effects of the response of
the RXTE photo-multiplier after high energy events, such as strong cosmic
ray showers. In the 90 minutes of RXTE data analyzed only $12$ of
the original $58$ candidates cannot be ruled out as artifacts, but
are hard to confirm as events \citep{2008ApJ...677.1241J,
 2007MNRAS.378.1287C, 2008MNRAS.388L..44L}. New RXTE/PCA data of Sco X-1 provided a less
constraining upper limit to the size distribution of KBOs
\citep{2008MNRAS.388L..44L}.
\begin{figure}[h!]
\centerline{\includegraphics[width=0.4\textwidth]{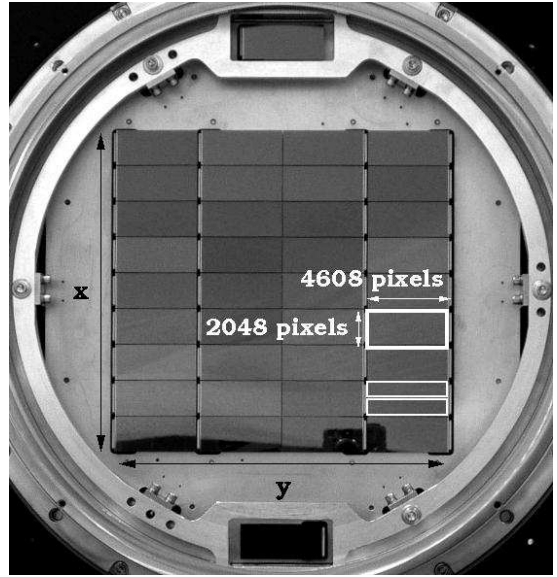}}
\caption[]{Megacam focal plane
 \citep{2006sda..conf..337M}. A thick rectangle outlines a
 single CCD in the 9x4 CCDs mosaic. Two halves of each CCD (thin
 rectangles) are read into two separate amplifiers; each amplifier
 generates a separate output image in our observational mode. The $x$
 and $y$ axis, as they would appear in a resulting image, are
 also shown.}
\label{fig:ccd}
\end{figure}

Several groups have conducted occultation surveys in the optical
regime. \citet{2006AJ....132..819R} and \citet{2008AJ....135.1039B}
independently observed narrow fields at $45~\mathrm{Hz}$ and
$40~\mathrm{Hz}$, respectively, with frame transfer cameras.  Such
cameras allowed them to obtain high signal-to-noise ratio (SNR) fast
photometry on two stars simultaneously. Both surveys expect a very low
event rate due to the limited number of stars and the limited
exposure, and neither survey has claimed any detection of objects in
the Kuiper Belt at this time\footnote{\citet{2006AJ....132..819R}
  report 3 possible occultations from objects outside of the Kuiper
  Belt.}.  An upper limit for KBOs with $d\ge 1 ~\mathrm{km}$ was
derived by \cite{2008AJ....135.1039B} by combining the non-detection
result of the surveys of \cite{2007MNRAS.378.1287C},
\citet{2006AJ....132..819R}, and \citet{2008AJ....135.1039B}. TAOS
(Taiwanese American Occultation Survey) is a dedicated automated
multi-telescope survey \citep{2008arXiv0802.0303L} that has observed a
set of fields comprising $\sim500$ target stars for over 3 years,
collecting over 150,000 star hours. TAOS reported no detections but
placed the strongest upper limit to date to the surface density of
small KBOs \citep{2008ApJ...685L.157Z}. TAOS observes at slower
cadence (4 or 5 Hz) and has a relatively low sensitivity
($\mathrm{SNR}\approx40$ at magnitude $R=12$). For these reasons TAOS
is only marginally sensitive to sub-km objects (with recovery
efficiency $\epsilon_{\mathrm{TAOS}}\approx0.3\%$ at 700 m and
$\epsilon_{\mathrm{TAOS}}\approx0.03\%$ at 500 m).

The survey we report here was conducted using Megacam
\citep[Figure~\ref{fig:ccd}]{2006sda..conf..337M} at the 6.5 m
MMT Observatory at Mount Hopkins,
Arizona. The use of Megacam in \emph{continuous--readout} mode (see
Section ~\ref{sec:contreadout}) on a field of view of $24'\times 24'$
allowed us to monitor over 100 stars at 200~Hz over the course of two
observational campaigns conducted in June 2007 and June-July 2008. Our
survey is sensitive to occultations of OSS objects $d
\sim700~\mathrm{m}$ or larger and we report no detections in 220 star
hours.  Our MMT survey is designed to be complementary to TAOS and to
reach smaller size limits, and unlike TAOS it would allow us to
estimate the size of a detected occulting KBO.  We expect further work
on adaptive photometry and de-trending to significantly improve our
sensitivity, perhaps allowing us to detect KBOs as small as
$d\ge300~\mathrm{m}$. We discuss the improvements we are developing on
this analysis in Section~\ref{sec:conc}. The preliminary analysis we
present here allows us to derive upper limits for objects
$d\geq700~\mathrm{m}$.

In the next section we describe the novel observational mode adopted
for this survey. In Section~\ref{sec:data} we describe the data
acquired and analyzed for this paper. Details of the data extraction
and reduction, which required custom packages, are addressed in the
same section. Section~\ref{sec:noise} describes the characteristics of
the noise of our current datasets, and our noise mitigation
approach. Section~\ref{sec:detection} describes the detection
algorithm. In Section~\ref{sec:limits} we derive our upper limit to
the density of KBOs. We also compare in detail the achievements of our
survey to those of previous surveys. We draw our conclusions and
outline future work in Section~\ref{sec:conc}.

\begin{figure}[ht!]
\centerline{\includegraphics[width=1.2\textwidth,
   angle=90]{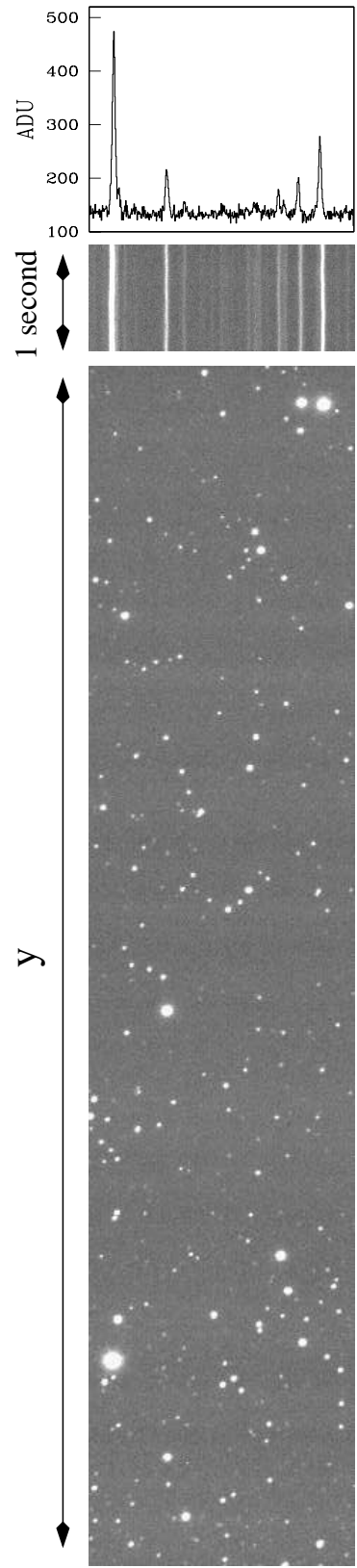}}
\caption[]{Conventional stare mode image (one half of a CCD) of one of
 our fields (bottom panel). A series of rows from continuous--readout
 mode (center panel) from the same CCD and field, where the rows are
 stacked together in a single image. The flux profile of the central row of
 this segment of continuous--readout data is plotted in the top panel.}
\label{fig:dataimg}
\end{figure}

\section{Fast Photometry with a  Large telescope: The Continuous--Readout Mode}\label{sec:contreadout}

\begin{figure}[bt]
\centerline{\includegraphics[width=0.5\textwidth]{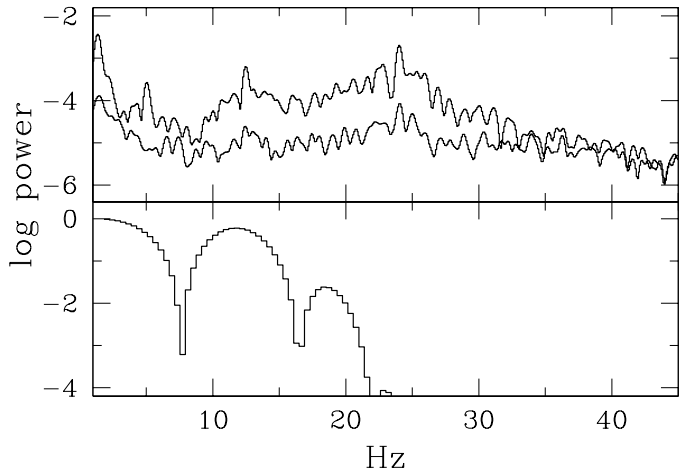}}
\caption[]{Top: power spectrum of one of our lightcurves before and after de-trending the lightcurve to remove noise (see Section \ref{sec:detrending}).
Bottom: power spectrum of the occultation time--series for a 1~km KBO  at 40 AU occulting a F0V $~V=~12$ star.}
\label{fig:fourier}
\end{figure}

Achieving sub-second photometric sampling is a challenge in optical
astronomy. CCD Cameras can perform fast photometric observations by
reading out small sub-images, limiting the
observations to very small portions of the sky (e.g.,
\citealt{2006AIPC..848..808M}).
This is the approach adopted by \citet{2006AJ....132..819R}, and
\citet{2008AJ....135.1039B}, who observed two stars at one time. 
Due to the rarity of occultation events, however, one would want to maximize the
number of targets and the total exposure to increase the number of detections. 
TAOS
achieves sub-second photometric observation on up to 500 targets with
the \emph{zipper mode} readout technique \citep{2008arXiv0802.0303L}, but they sample at
$\leq~5~\mathrm{Hz}$ rate.  Our continuous--readout technique allows us to
observe the entire field of view of the camera at $200~\mathrm{Hz}$.

Megacam, the MMT optical imager, is a mosaic
camera comprising thirty-six CCDs --~each with an array of $2048
\times 4608$ pixels --~with a $24' \times 24'$ field of view
(Figure~\ref{fig:ccd}). The standard readout speed of each CCD is
0.005~sec/row with $2 \times 2$ binning. For this survey, we operated
the camera in shutterless continuous--readout mode; that is, we kept
the shutter open while scrolling and reading the charges at the
standard readout speed, tracking the sky at the sidereal rate. Each
star is represented in each row that is read out of the camera, and
the flux from a star in a row represents a photometric measurement of
that star sampled at 200~Hz. Stacking each read row into a single
image each star time--series forms a streak along the readout axis
($y\mathrm{-axis}$).  A small portion of our data is shown in
Figure~\ref{fig:dataimg}.

In this observational mode the flux from the sky background is added
continuously as the charge is transferred from one end of the CCD to
the other, so the sky is exposed $2304\times 0.005 ~=~ 11.52
~\mathrm{sec}$ for every 0.005 sec integration on each star image
(where 2304 is the effective number of rows in each $2\times 2$ binned
CCD). In this mode the photon limited SNR is typically $\sim180$ for an $r'$
magnitude 10 star.

When observing multiple targets simultaneously one can notice
that
the star
lightcurves are affected by common fluctuations, or \emph{trends}, due
for example to weather patterns \citep[and references
therein]{2008arXiv0812.1010K}.  In our observational mode, however,
additional flux variations are caused by wind-induced resonant
oscillations of the telescope.  While the image motion along the $x$
axis of the focal plane (transverse to the readout direction) can be resolved
(see Section \ref{sec:extraction}), the image motion parallel to the
direction of the CCD readout induces an effective variation in the
exposure time of a star for a given row.  These fluctuations are
common to all stars in the field (with possible position dependencies)
and therefore, in principle, they are completely removable. We discuss
the de-trending of our data in Section~\ref{sec:detrending}.  Other
sources of noise that affect continuous-readout mode data are
discussed in Section~\ref{sec:noise}. The typical duration of a set of
contiguous data was 10--15 minutes (after which the data load
on the buffer would become prohibitive).  For each amplifier, a single
FITS\footnote{Flexible Image Transport System,
\url{http://fits.gsfc.nasa.gov/}.} file is created wherein all of the
rows read out during a data run are stored as a single image.  For a
typical run each FITS output image contains $100\mathrm{K}$ to
$130\mathrm{K}$ rows, corresponding to about 150--200~Mb of data.

\section{Data}\label{sec:data}

\begin{deluxetable}{cccc}
\tablewidth{0pc}
\tablecaption{Parameters of the observed fields}
\tablehead{RA & Dec & $\lambda$\tablenotemark{a} & $\varepsilon$ range\tablenotemark{b} \\
 & & (deg) & (deg)} \startdata
$17\h 00\m 00\s$ & $-21\degree 15' 00''$  &  1.5 & 174--160\\
$17\h 15\m 00\s$ & $-20\degree 15' 00''$  & 2.8 & 176--163\\
$18\h 00\m 00\s$ & $-21\degree 15' 00''$  & 2.2 & 171--173\\
$18\h 00\m 00\s$ & $-21\degree 30' 00''$  & 1.9 & 171--173\\
$18\h 00\m 00\s$ & $-21\degree 45' 00''$  & 1.7 & 172--173\\
$19\h 00\m 00\s$ & $-22\degree 00' 00''$ & 0.7 & 158--172\\\enddata
\tablenotetext{a}{ecliptic longitude}
\tablenotetext{b}{range of elongation angles}
\label{tab:pointing}
\end{deluxetable}
We selected observing fields within $2.8\degree$ of the ecliptic
plane, where the concentration of KBOs is highest
\citep{2001AJ....121.2804B}. In order to maximize the number of
targets we selected our fields at the intersection of the ecliptic and
galactic planes ($\mathrm{RA}~\sim 19\h 00\m 00\s$,
$~\mathrm{Dec}~\sim -21^o 00' 00''$). We conducted our observations in
June-July, when our fields were near opposition (elongation angle
$\varepsilon~=~180\degree$) and the relative velocity of the KBOs is
highest \citep{1987AJ.....93.1549R, 2007AJ....134.1596N,
  2009arXiv0902.3457B}, thus maximizing the event rate per target
star. Pointing information for our fields is summarized in Table
\ref{tab:pointing}. The RA and Dec of each
observed field are listed together with the ecliptic latitude
($\lambda$) and a maximum range of elongation angles at which the
filed might have been observed.

We also observed control fields.  These were chosen on the galactic
plane at a high ecliptic latitude; we expect a negligible rate of
occultations by KBOs in these fields.  These data allow us to assess
our false positive rate. Since we report no detections the analysis of
these fields is not discussed further in this paper. All of our
observations were conducted in Sloan $r'$ filter
\citep{2002AJ....123.2121S}. A set of about $7$ hours on target fields
was collected in 5 half nights in June 2007 and a similar number of
hours was collected on control fields. A set of about $7$ hours on
target fields and about $6$ hours on control fields was collected
in 7 half nights in June-July 2008.  Out of the 2007 dataset 100.61
star hours at $\mathrm{SNR} \ge 25$ are considered in this paper.
From the 2008 dataset we use here 118.93 star hours.  Information on
our dataset is summarized in Table \ref{tab:dataset}.  The minimum
signal-to-noise ratio of 25, is chosen arbitrarily: 25 is the minimum
$\mathrm{SNR}$ of the surveys of \citet{2006AJ....132..819R} and
\citet{2008AJ....135.1039B}.\footnote{Note however that this SNR level
  is obtained here for 200 Hz, whereas \citet{2006AJ....132..819R} and
  \citet{2008AJ....135.1039B} observed at 45 Hz and 40 Hz.} A SNR 25
limits our sensitivity to fluctuations greater than $4\%$. An
occultation of a magnitude 12 F0V star by a KBO of $d = 400~\mathrm{m}$
diameter would produce a $4\%$ effect. Our efficiency tests, however,
revealed our sensitivity rapidly drops below 10\% for objects smaller
than $d=700~\mathrm{m}$, due to residual non-Gaussianity in our
time--series photometric data. We discuss this in
Section~\ref{sec:noise}.

\begin{deluxetable}{ll}
 \tablewidth{0pc}
\tablecaption{Data set parameters.}
\tablehead{ }\startdata Start Date & 2007 June 6\\ End Date & 2007
June 10\\ Exposure at
SN$\ge25$ & 100.61 star-hours\\ Number of lightcurves with SN$\ge25$ &
990\\ Number of Photometric Measurements& $ 7.2\times10^7$\\ \hline Start Date & 2008 June 27\\ End Date & 2008 July 1
\\ Exposure at SNR$\ge25$ & 118.93
star-hours\\ Number of lightcurves with SNR$\ge25$ & 527\\ Number of
Photometric Measurements & $ 8.5\times10^7$\enddata
\label{tab:dataset}
\end{deluxetable}

\subsection{Data extraction and reduction}
\subsubsection{Extraction}\label{sec:extraction}
Custom algorithms have been developed for the data extraction and
reduction.  For each field a preliminary \emph{stare mode}
(conventional) image is collected before each series of high-speed
runs. At the beginning of our analysis the stare mode image is
analyzed using SExtractor \citep{sextractor} to generate a catalog of
bright sources. This catalog is used to identify the initial position
and brightness of each star in the focal pane.  In order to analyze
the continuous readout data, we first determine the sky background for
each CCD and each row.  To do so we calculate the mean of the flux
counts in each row after removing the measurements that
are three $\sigma$'s or more from the mean ($3\sigma$-clipping)
iteratively until the mean converges.  This removes most of the pixels in the row
containing flux from resolved stars.  Next, a subset of stars that are
bright and isolated is selected from the stare--mode catalog and used
to determine the $x$-displacement of the focal plane.  The focal plane
is split into two halves, $9\times2$ chips each, that are analyzed
separately. We select eight stars, two near each of the four corners
of each half-focal plane.  This allows us to characterize the global
motion of the targets even in the presence of small rotational modes
or spatial dependency (see Section~\ref{sec:noise}).  For each star
($\star$), and at each time--stamp (t), we calculate $\mu_\star(t)$
and $\sigma_\star(t)$, respectively the centroid offset from the
original position and the standard deviation of the star image,
assuming a Gaussian profile. Note that, for a given time-stamp, flux
from different stars will appear on different rows due to the
$y$-positions of the stars on the focal plane. A 1-D Gaussian
\begin{equation}
F_\star~=~I_\star~\exp\left(-\frac{(x-\mu_\star(t))^2}{2\sigma_\star^2(t)}\right) ~+
~I_{bg}
\end{equation}
(where $F_\star$ is the total star flux, $I_\star$ the flux at the
peak and $I_{\mathrm{bg}}$ the sky) is fit for each of the eight stars
to each row of the star--streak.  Thus the $x$-displacement
$\bar{\mu}(t)$ for all the stars in the field at time--stamp $t$ is
estimated to be the weighted average of the star displacements:
 \begin{equation}
\bar{\mu}(t)~=~\frac{\sum\limits_{\star=1}^8
 \omega_\star(\mu_\star(t)-\mu_\star(t_0))}{\sum\limits_{\star=1}^8
 \omega_\star},
\end{equation}
where $\mu_\star(t_0)$ is the star initial $x$-position and
$\omega_\star$ is the weight used for that star.

In order to weight our average
we use the correlation of the entire $x$--displacement time--series
${\mathbf \mu_\star}$ with respect to the rest of the star set:
\begin{eqnarray}
\omega{(i,j)} &=& \frac{1}{T}\sum_{t=0}^T
\frac{(\mu_i(t)-\left<\mu_i\right>)(\mu_j(t)-\left<\mu_j\right>)}{s_i^2(t)~s_j^2(t)},\\
\omega_\star  &=& \frac{1}{7}\sum_{j\neq \star} \omega(\star,j);
\end{eqnarray}
where $s^2$ is the variance of the displacement throughout the duration $T$ of the time--series. The weight $\omega_\star$ is the square of the Pearson's
correlation coefficient \citep[pag. 406]{rice2001}, a measure of the correlation of the displacement time--series for one star with the other seven.
 All star
lightcurves in the field are then extracted by aperture photometry
adjusting time--stamp by time--stamp the center of the aperture
according to the $x$-motion derived in this stage, and with a fixed
aperture size which is proportional to the average FWHM in the
run.\footnote{We attempted to extract the lightcurves with both fixed
 aperture size and variable aperture size, using the FWHM calculated
 by Gaussian fitting as a point by point estimator of the aperture
 size.  The fixed aperture extraction proved to be more reliable than
 the variable aperture extraction, which induced further
 noise in our lightcurves.}

\subsubsection{De-trending}\label{sec:detrending}
The lightcurves thus extracted show evident semi--periodic,
quasi--sinusoidal flux variations that can be associated with
oscillatory modes of the telescope in the $y$ direction.  In
particular, a Fourier analysis generally reveals two strong modes,
roughly consistent among runs, one with period near $0.04$ seconds and
the other near $0.5$ seconds. Fourier spectra for one of our lightcurves,
before and after processing it, are shown in Figure~\ref{fig:fourier}
(top).  Because these fluctuations affect the whole CCD plane, they
are common to all stars and can be removed to achieve greater
photometric precision. We now want to identify and remove these trends
from our lightcurves, a process that we call \emph{de-trending}.

\begin{figure*}[bt]
\centerline{\includegraphics[width=0.5\textwidth]{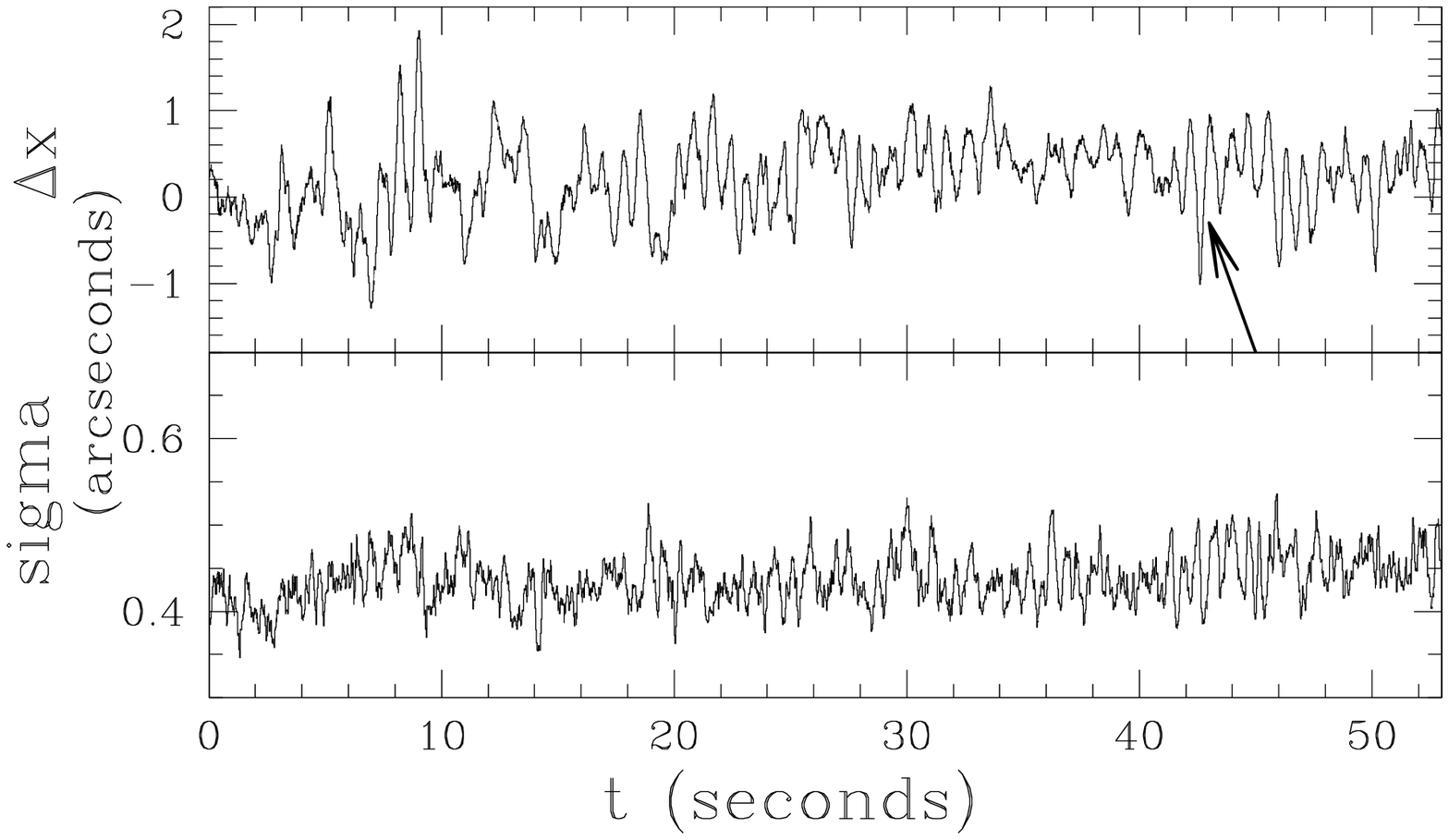}
\includegraphics[width=0.5\textwidth]{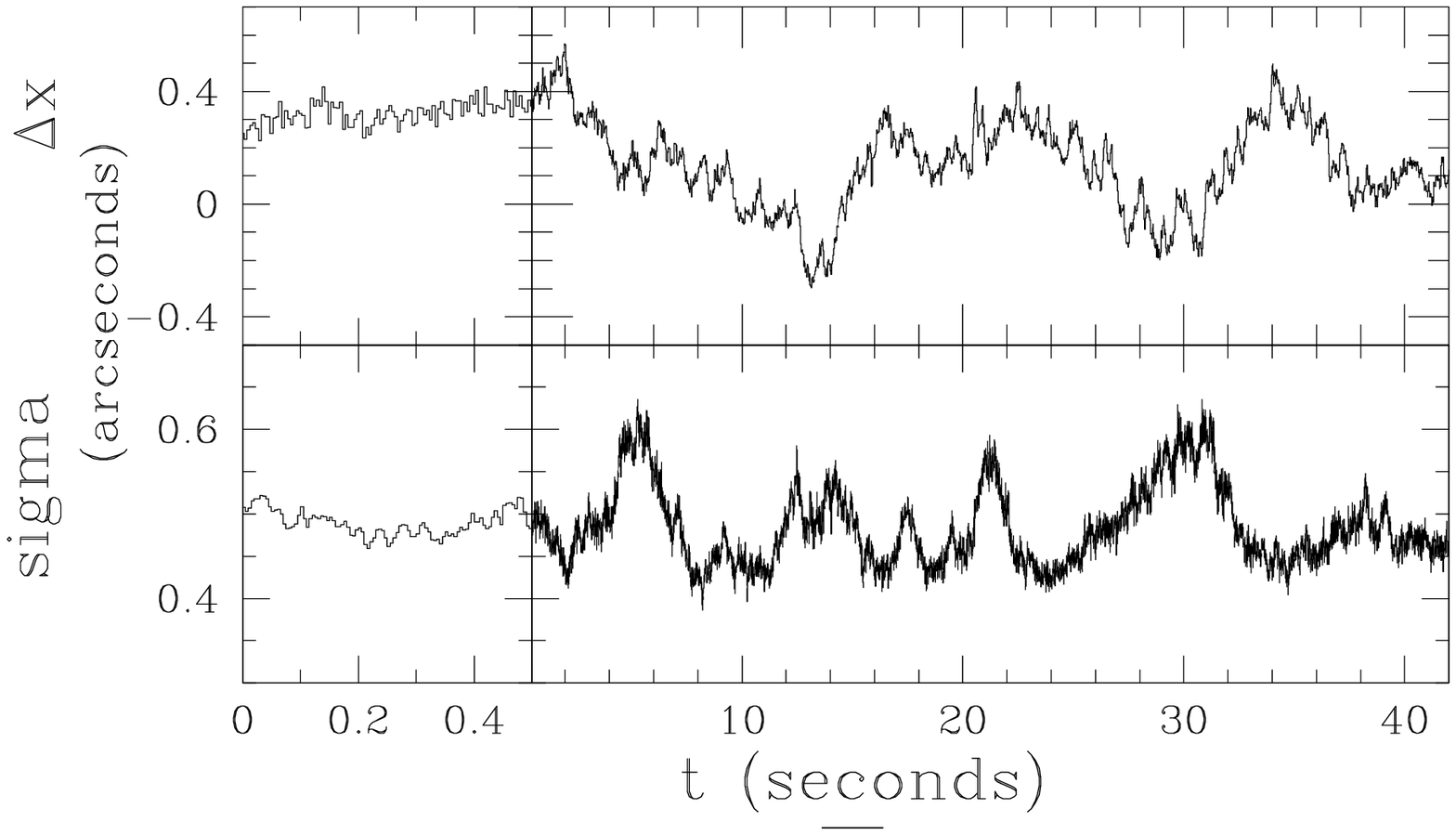}}
\caption[]{Image motion and PSF over time: mean of the $x$ displacements  for eight bright isolated stars, at the four corners of the half-focal plane for two data runs (top left and right panels).  PSF width from the Gaussian fit averaged over the same set of stars (bottom left and right).
On the left we used the same run used to generate Figure~\ref{fig:streaks}. The arrow points to the displacement feature marked in Figure \ref{fig:streaks}.
On the right the $x$-displacement and the PSF width for another run, with the first 0.5 seconds shown on the left at higher time resolution. Note how in the second run the $x$-displacements are less prominent (note the different $y$ scale) but the amplitude of the variability of the PSF is larger.}
\label{fig:xmotion}
\end{figure*}

The general algorithm we used for de-trending is described in
\cite{2008arXiv0812.1010K}. The method takes advantage of the
correlation among lightcurves to extract and remove common
features.  Since we can identify distinct semi--periodic modes we
de-trend high and low frequencies separately (typically
$\nu>10\mathrm{Hz}$ and $\nu<10\mathrm{Hz}$).

We first smooth the lightcurves, to remove all but the frequencies that
we want to de-trend, by applying a low--pass or high--pass filter.  We
then select a subset of $N_\tau$ \emph{template} lightcurves
($f_\tau$) that show the highest correlation in the lightcurve
features. $N_\tau$ is typically about 15.  A master trend lightcurve
$\tau$ is generated as the weighted average of the normalized template
lightcurves:
\begin{equation}
\tau(t) = \frac{1}{N_\tau}~\frac{\sum\limits_{j=1}^{N_\tau} \sigma^2(f_{\tau,j})~f_{\tau, j}(t)/\langle f_{\tau, j}\rangle}{\sum\limits_{j=1}^{N_\tau} \sigma^2(f_{\tau,j})}
\end{equation}
where the notation $\langle f_{\tau, j}\rangle$ denotes the mean flux of $f_{\tau, j}(t)$ over the duration $T$ of the
lightcurve, and the weight $ \sigma^2(f_{\tau,j})$ is the variance of the lightcurve in time;
$\tau(t)$ has mean value of unity and it represents the correlated fluctuations in all lightcurves.

The main trend is physically associated with an over-under exposure
phenomenon due to global image motion along the $y$ axis, which causes
the effective exposure time to vary (see
Section~\ref{sec:contreadout}), therefore scaling the flux.  In order
to remove these common trends we divide point by point the
flux of each original lightcurve $f$ by the trend master
lightcurve. To improve the de-trending effectiveness we allow a free
multiplicative factor $A_f$ (a scaling factor) for each lightcurve as
follows: \begin{equation} f_{\mathrm{d},
A_f}(t)~=~f(t) \left[\left(\frac{1}{\tau(t)}-1\right)A_f~+~1\right];
\end{equation}
$f_{\mathrm{d}, A_f}$ is the de-trended lightcurve.

We
optimize our de-trending by selecting $A_f$ to minimize the variance of the
de-trended lightcurve $f_\mathrm{d}$ with respect to
$f_\mathrm{c}~=~f-f_\mathrm{s}+\langle f_s\rangle$,
which is the original lightcurve cleaned of the frequency to be
de-trended. We apply a high--pass (low--pass) filter to $f$ to obtain $f_\mathrm{s}$ if we want to de-trend the low (high) frequencies.
$A_f$
is then optimized by setting:
\begin{equation}\frac{\partial}{\partial A_f} \sum_{t=1}^T
\left(f_{\mathrm{d}, A_f}(t)~ - ~ \ave{f_\mathrm{c}}\right)^2 =
0, \end{equation}
which minimizes the
second moment of the de-trended lightcurve with respect to
$f_\mathrm{c}$. The optimal value of $A_f$ can be calculated analytically.

We set no constraints on $A_f$, and for all of our runs the optimal values of $A_f$ proved to be close to 1 (which is what we expect in the presence of global trends) except for pathological cases where the flux of the star was buried in noise and the raw and detrended SNR were extremely low. These lightcurves would not pass SNR cuts and were never considered in any of our analysis.

\begin{figure}[bt]
\centerline{\includegraphics[width=0.5\textwidth]{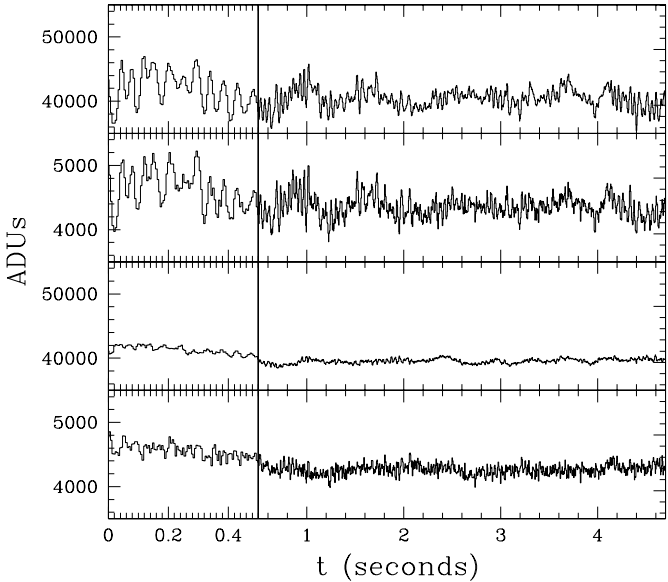}}
\caption[]{Lightcurves of two independent stars in one of our fields. The left-hand plots show a 0.5-second chunk of the time series; the following 4-seconds  are shown on the right at a lower time resolution.
The top two panels show the lightcurves before detrending. Common modes are visible at multiple time scales.
The bottom two panels show the lightcurves after detrending.  The top lightcurve is the same used in Figure~\ref{fig:fourier}}
\label{fig:detrending}
\end{figure}

\begin{figure}[bt]
\centerline{\includegraphics[width=0.5\textwidth]{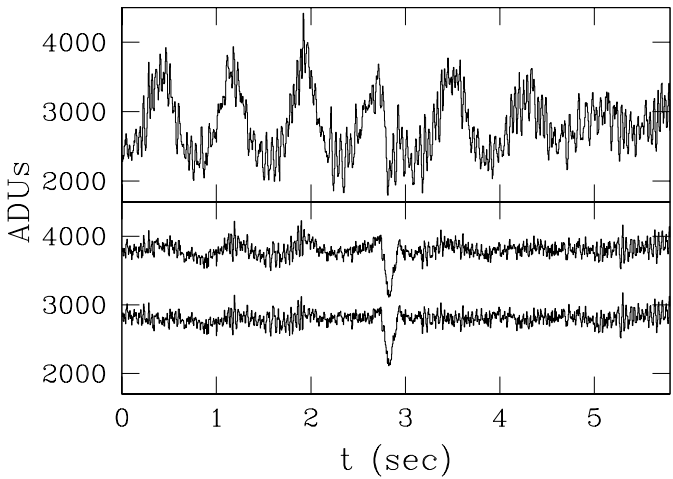}}
\caption[]{Raw lightcurve on which the occultation signature of a 1~km
 KBO occulting a  magnitude $V~=~9$ F0V star has been implanted (top)
 and the same lightcurve after detrending (bottom). In the bottom
 panel the top lightcurve is detrended without optimization (plotted
 at the top at an arbitrary offset) and the bottom lightcurve is
 detrended with optimization factor $A_f~=~1.15$ for the low
 frequencies and $1.05$ for the high frequencies.}
\label{fig:optimized}
\end{figure}

Examples of the results obtained by our de-trending algorithm are
displayed in Figures~\ref{fig:detrending} and \ref{fig:optimized}. In
Figure~\ref{fig:detrending} the top two panels show lightcurves for
two independent sources in our field, and the bottom two panels show
the same lightcurves after de-trending. Note that the top star is $\sim2.5$
magnitudes brighter than the other and this is reflected in the lower
SNR of the fainter source (bottom panel). Figure~\ref{fig:optimized}
shows one of our lightcurve before (top) and after de-trending
(bottom). The raw lightcurve is implanted with an occultation by a
$d~=~1~\mkm$ KBO occulting a $V=9$ F0V star. The diffraction feature
is completely lost in the trends and becomes evident only after
de-trending. In the bottom panel we show the lightcurve detrended
without allowing for the optimization factor $A_f$ at the top (plotted
at the top at an arbitrary offset) and with optimization factor
$A_f~=~1.15$ for the low frequencies and $A_f~=~1.05$ for the high
frequencies, shown at the bottom. The introduction of an optimization
factor improves the SNR of the de-trended lightcurve from
$\mathrm{SNR}~=~30.0$ to $\mathrm{SNR}~=~30.7$. For this particular
run improvements of up to 7\% in SNR were achieved by optimizing the
detrending.

Note that, while we used smoothed versions of our lightcurves to
identify the trends and to optimize the de-trending, we do not smooth
or filter our lightcurves to improve the SNR, thus preserving all
intrinsic features (including potential occultations).
Figure~\ref{fig:fourier} shows the power spectrum of one of our
lightcurves before and after de-trending it (top). The power spectrum
of an occultation time-series generated by a 1~km KBO occulting a F0V
star of magnitude $V~=~12$ is shown in the bottom panel. Our
de-trending greatly reduced the power at all frequencies: the
cumulative power for this particular lightcurve at frequencies
$\nu\leq 40~\mathrm{Hz}$ is suppressed by a factor of 40.
Because the
oscillations are not perfectly correlated among our stars (see
Section~\ref{sec:noise}) some residual power is visible. Smoothing
however would would significantly reduce the strength of the
occultation features, that show power at all frequencies
$\nu<20~\mathrm{Hz}$.

\section{Residual noise in the time-series}\label{sec:noise}
\begin{figure}[bt]
\centerline{\includegraphics[width=0.8\textwidth,
   angle=90]{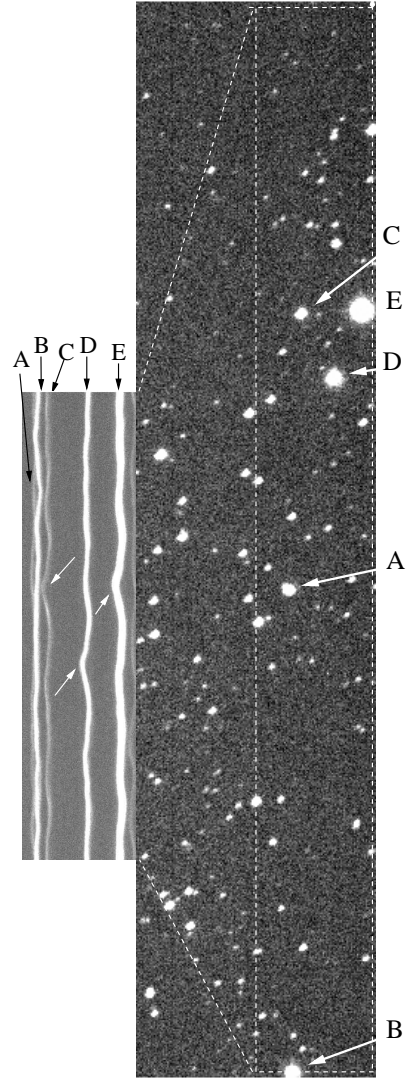}}
\caption[]{A stare mode image (right) read out from a single amplifier and a corresponding $\sim10$~second chunk of high frequency data (left). A few bright stars and the corresponding streaks are indicated by letters. White arrows in the left panel point to a distinctive synchronous displacement feature in the data, visible clearly in three of the streaks, in order to focus the reader's attention to the non-parallelism of simultaneous features in our data, which is due to offset in the original $y$ position.}
\label{fig:streaks}
\end{figure}

With a SNR $\gtrsim25$ we can detect fluctuations of a few--percent. In an 0.005~sec exposure the flux for a magnitude
$r'=14$ star observed by Megacam is about $10^3~e^-$, which after taking into account the contribution to the noise of the background should lead to a Poisson limited SNR of about $25$.
While we were able to remove a large portion of the noise that originally affected our data, we typically cannot reach the Poisson-limit.
We have identified five possible sources of noise in our data:
\begin{itemize}
\item {\bf Contamination by nearby sources}.  Overlap of stars along
 the $x$ axis (perpendicular to the read-out direction) within a
 chip, causes reciprocal contamination in our readout
 mode and some of the stars in our fields are therefore compromised
 and excluded from our analysis.  Furthermore, oscillations of the
 images along the $x$ axis causes the relative distance between the
 star--streaks to change, which causes occasional merging. Note that
 while these oscillations are simultaneous in time domain, they do not occur in the same row in the recorded image. In each row the star images
 of two objects that are at a different $y$ position on the
 CCD plane will not belong to the same time-stamp, therefore the
 oscillations --~while simultaneous~-- will show a $y$ offset.  This is
 shown in Figure~\ref{fig:streaks}. The merging of streaks
 causes artificially high counts. Aperture photometry with a fixed
 aperture does not address this issue properly and fitting photometry
 on individual streaks is a computationally expensive, inefficient
 method which is also unstable in the presence of multiple sources
 close to each other.

\item {\bf Unresolved sources}. Sources that are too faint to be
  visible in our 0.005 sec exposures generate a diffuse background.
  For the data in Figure~\ref{fig:dataimg} the sky level calculated as
  the $3\sigma\mathrm{-clipped}$ mean of the row counts is 140.5
  ADUs. The stare mode image sky level was 48 ADUs for a 5 sec
  exposure, which would lead to a prediction of 110 ADUs for our
  $0.005\times 2304$ sec effective continuous--readout exposure. The
  discrepancy is due to the presence of unresolved streaks associated
  with faint stars across the field. Summing all the counts in the
  stare mode image and rescaling by the exposure time of each row we
  get a number very close to the sum of all counts in a row of
  continuous-readout data.  This contamination introduces extra
  Poisson noise, but more importantly it introduces non-Poissonian
  noise as well, since the unresolved sources are affected by the same
  trends the bright stars display. Our data shows evidence of
  off--phase correlation that might be induced by unresolved sources.

\item {\bf Positional dependency in the motion and trends}.  While we
 treat all of the stars in the field as an ensemble that moves in a
 solid fashion along the $x$ and $y$ axes, the image motion might
 also have a rotational component. This would lead to position
 dependencies in the motion that are not accounted for by our
 aperture centering algorithm. We have not seen evidence of
 dependency on the distance to the center of the focal plane in
 either motion or trends, but we cannot exclude that occasional
 rotational modes of the telescope would occur. Differential image
 motion and flux fluctuations might also be induced by atmospheric
 seeing. Both of these effects might cause the star--streak to move
 out of the photometric aperture leading to artificially low counts.
 The aperture size must be chosen to be such that errors due to
 contamination by nearby sources and errors due to streaks exiting
 the aperture are simultaneously minimized. Furthermore in the
 presence of thin clouds, variations in the transparency might generate
 trends that would affect different sources at different times as the
 clouds move across the image. Positional dependencies or variations
 in transparency might contribute to the off--phase correlation of
 our data.

\item{\bf{Scintillation}}. 
Young's  scaling law (\citealt{1967AJ.....72..747Y}, \citealt{1993AJ....106.2441G}, \citealt{1998PASP..110..610D}),
\begin{equation}
\sigma_I ~=~ \frac{0.09A^{-2/3}(\sec Z)^{1.75}exp(-\frac{h}{h_0})}{\sqrt{2~\Delta t}},
\end{equation}
describes the error in flux intensity $I$ due to the low--frequency component of scintillation, with $\sigma = \left(\Delta I/I\right)$ and where $A$ is the telescope aperture (in cm), $Z$ the angle from zenith, $h$ is the height of the turbulence layer, the scaling factor $h_0 = 8000~\mathrm{m}$,  and $\Delta T$ the integration time. Competing effects are in place in our survey: the large aperture mitigating the noise, and the low air mass contributing to signal degradation. Note however that this relation holds for integration time on scales of seconds or longer.
When including the effects of high--frequency scintillation the dependency on the aperture is expected to be steeper:
\begin{equation}
\sigma_I^2 \propto A^{-7/3}(\sec Z)^3\int_0^{\infty} C_n^2(h)h^2d h,
\end{equation}
where $C_n$ is the refraction
coefficient for the turbulent layer (see \citealt{1997PASP..109..173D} and references therein).

Using the above equations and representative data from La Palma \citep{1998PASP..110..610D} we estimate that the noise contribution from scintillation is $\sigma_I < 0.01$, i.e., not the dominant source of residual noise. 
As compared to the other occultation surveys the
term associated to the telescope aperture in the SNR
variance ($A^{-7/6}$)  is a factor 20 lower than the same factor for the TAOS
survey, 4.5 times lower than the same factor for
\citet{2008AJ....135.1039B} and 1.4 than for the
\citet{2006AJ....132..819R} survey.

\item {\bf Convolution of the time series with finite PSF}. The finite
 size of the PSF (typically two to three pixels, although it
 occasionally was as large as seven) causes consecutive measurements
 to be correlated. This effect is not a source of noise \emph{per
   se}, but it changes the spectral characteristics of the noise. The
 scale of this phenomenon shows up in
 an autocorrelation analysis with 
 high power at a lag of about seven pixels. This is effectively a kernel
 convolution of our time series that smoothes the signal, including
 possible occultation signals, so that while we sample
 the images at 200 Hz we 
 would expect an occultation signature to be effectively sampled at $\approx30~\mathrm{Hz}$
 (see Section~\ref{sec:detection}). Note that this is close to, but
 slightly short of, the critical Nyquist sampling for occultations
 dominated by diffraction \citep{2009arXiv0902.3457B}.

\end{itemize}

While we achieved significant noise reduction with our detrending, our
SNR is typically a factor of two to three lower than the
Poisson--limit. Our noise is characterized by high kurtosis, which is
indicative of non--Gaussianity. Residual low frequency fluctuations
(about 100 points period) are still noticeable in many of our time
series (see Figure \ref{fig:detrending}). Possible improvements are
discussed in Section~\ref{sec:conc}.

 \begin{figure}[bt]
\centerline{\includegraphics[width=0.5\textwidth]{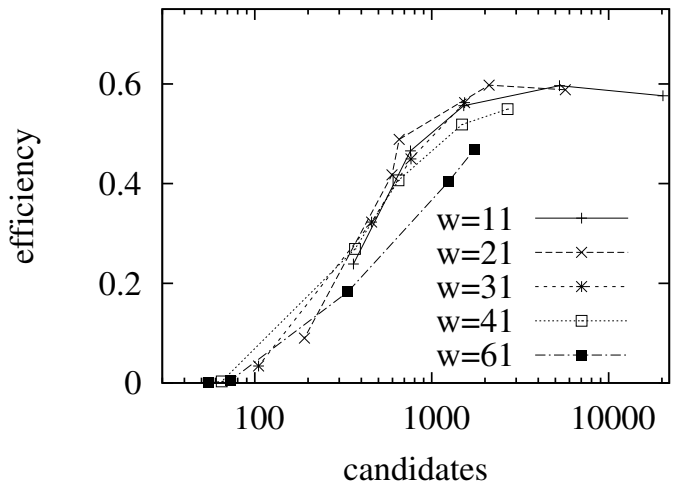}}
\caption[]{Efficiency plotted against the number of unidentified  candidates (mostly false positives). Each line represents a different window size $w$, and each point represents the value of the efficiency at threshold $\theta~=$ 0.08, 0.10, 0.15, 0.20, 0.30, the number of false positives monotonically grows with decreasing $\theta$ (larger values of $\theta$ on the left). All lines (all $w$ values) show a plateau at different thresholds.}
\label{fig:roc}
\end{figure}

\section{Search for events and efficiency}\label{sec:detection}
\subsection{Detection algorithm}
The signature of an occultation, sampled at any rate $\gtrsim
20~\mathrm{Hz}$, is very distinctive: it shows several fluctuations
prior to the Airy ring peak, then a deep trough and possibly a Poisson
spot feature, followed by a second Airy ring rise and more
fluctuations (see Figure~\ref{fig:occultationlcv}). The prominence of
these features depends upon the magnitude and spectral type of the
background star, which together determine the angular size, as well as the size
and the sphericity of the occulter, distance to the occulter, and
impact parameter \citep{2007AJ....134.1596N}.

One possible approach to detecting occultations in our lightcurves is to
take advantage of this peculiar shape, for example using correlation
of templates, as in \cite{2008AJ....135.1039B}. Given the
size of our dataset, however, we chose to utilize a search algorithm general
enough to capture any fluctuation of some significance, but which
requires less computational power.  We scan our time series for any
fluctuation lasting longer than a duration $w$, and on average
greater than a threshold $\theta$ from the local mean, which is
calculated over a window $W$ of 300 data points surrounding $w$.
Windows $w$ of 11, 21, 31, 41 and 61 points were considered, in combination
with thresholds of 0.10, 0.15, 0.20, 0.30.
We define the local intensity $I_\mathrm{l}(i)$ as the ratio of the flux in the local window $w$
and in the surrounding window $W$.  If the flux in $w$ is suppressed by
more than our threshold $\theta$ from the local mean (mean over $W$),
\begin{equation}
I_\mathrm{l}(i)~=~\frac{\displaystyle\sum\limits_{j=i-w/2}^{i+w/2} f_j/w}{\displaystyle\sum\limits_{j=i+W/2}^{i+W/2} f_j/W}\leq1~-~\theta,
\end{equation}
then $w$ is considered as a candidate.  This is similar to the
Equivalent Width algorithm, which is used in spectral analysis, and
for rare event searches by \citet{2006AJ....132..819R} and
\citet{TAOS_Sedna}.  Overlapping candidates are then removed and the
center of the window $w$ that displayed the largest deviation is
selected as a single candidate event. Note that this algorithm would
in most cases trigger two separate events for the two halves of an
occultation on opposite sides of the Poisson spot
(Figure~\ref{fig:occultationlcv}). These cases are later automatically
recognized and accounted for as a single event. Different
choices of $w$ and $\theta$ will produce different detection
efficiency and false positive rate. We select an optimized subset of
combinations of $w$ and $\theta$ to be used for our event detection. This
optimization is described in the next section.

\subsection{Efficiency}
We test the efficiency of our search by implanting simulated
occultations in our raw lightcurves. By using our true dataset instead
of generating synthetic data we do not introduce any assumptions about
the nature of our time series.  We run the implanted lightcurves
through the same pipeline as the original lightcurves: de-trending
them and searching for significant deviations from the mean flux.  In
order to achieve better sampling of our efficiency the entire dataset
was implanted with one occultation per lightcurve at each KBO size we
tested: $d = 0.5, 0.6, 0.7, 0.8, 0.9, 1.0, 1.3, 2.0, 3.0~ \mkm$, and
the efficiency was assessed for each size separately. The finite PSF
width of the star induces correlations among consecutive time
stamps. Given the typical PSF size in our data (see
Figure~\ref{fig:xmotion}) measurements are considered independent if
separated by more than about seven pixels. Therefore, to modulate the
original time series by the occultation signal we multiply the star
flux by a synthetic occultation lightcurve sampled at
$30~\mathrm{Hz}$.\footnote{Since the occultation typically
  suppresses the flux, multiplying by the occultation signal reduces
  the noise by a factor proportional to the occultation flux decrease,
  causing us to overly suppress the Poisson noise by a factor of the
  square root of the modulation. Furthermore, sources of noise that
  are not proportional to the photon counts (such as sky background
  and read noise) should remain constant during an occultation event,
  but this noise is reduced by a factor of the flux modulation when
  the event is added to the lightcurve in this way. However, since we
  expect to have a very high recovery efficiency for any occultation
  which generates effects $ \geq 20\%$, where the underestimation would
  become significant, we do not expect this effect to impact our
  efficiency estimation.}

For the purpose of our efficiency simulations we assume all objects
are at 40 AU, since we expect our occultations to be within the Kuiper
Belt. There is little difference in the diffraction feature between 35
and 50~AU.  The differences in spectral power between the star types
do not impact the occultation features as observed by our system, so
we simulate all of our occultations assuming an F0V type star.  The angular size
of the star affects the shape of the occultation by smoothing the
diffraction features. It is therefore important to properly sample the
angular size space. We find that, given the objects in our fields,
imposing a flat prior to the magnitude distribution between $V~=~8$ and
$V~=~11$ adequately samples our angular size range.
The flat prior slightly overestimates the average cross section $H$ of the
events, but this effect is more than compensated by the loss in
efficiency due to the fact that, for stars with larger angular sizes,
the occultation signal is smoothed out as the diffraction pattern is
averaged over the surface of the star, making the event harder to
detect \citep{2007AJ....134.1596N}. Overall our estimate of our
detection rate is conservative.

To characterize our efficiency we implant occultations at random
impact parameters $b \in [0,H(d)/2]$. However, we first want to choose
the most appropriate window size and threshold combinations, and for
that we implant occultations by $d=1~\mathrm{km}$ KBOs in the reduced
impact parameter space $b~\in~ [0,~0.3\cdot H(d)]$. This set of modified
lightcurves is used to optimize our parameters to maximize our
efficiency and minimize the number of false positives simultaneously.
Although our generic detection approach can reach high efficiency
(nearly $100\%$ for $1~\mathrm{km}$ KBOs at zero impact parameter), it
also produces a large number of candidates, most of which are expected
to be false positives. The combination of $w$ and $\theta$ values
generated efficiencies ranging between 94\% (at $w$=11 and $\theta$ =
0.1) and 0 (at $w$=61 and $\theta$ = 0.3) and the number of candidates
ranged between 0 and over 1000.  Figure~\ref{fig:roc} shows the
behavior of the efficiency as a function of number of
candidates. Different window sizes are represented by different lines
and the different thresholds are marked by the points along each
line. Typically, after a rapid increase in efficiency with the
decreasing threshold, the efficiency stabilizes, while the number of
candidates
keeps  growing: we want to choose our parameters near this
point, where the efficiency is highest and any less stringent choice
would only increase the number of our false positives.  We select
combinations of $w$ and $\theta$ that yield both an efficiency $>50\%$
and a ratio of efficiency to candidates $<0.5$.
The following are the accepted windows-threshold
combinations: ($w$, $\theta$) = (21, 0.15), (31, 0.20), and (11, 0.25).
Events found in any run with these selection parameters were
considered as candidates. We reached an overall efficiency of 82\%
at $d= 1 ~\mathrm{km}$ for
lightcurves implanted with synthetic occultations at varying impact parameters
between 0 and $H/2$.

\begin{figure}[bt]
\centerline{\includegraphics[width=0.5\textwidth]{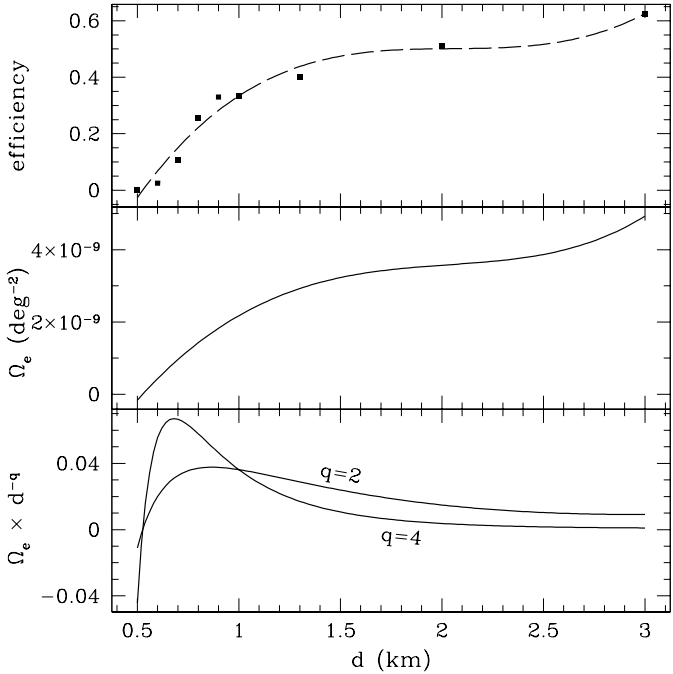}}
\caption[]{Efficiency of our survey as a function of
 KBO diameter (\emph{top}). The \emph{central} panel shows the
 effective solid angle of our survey and the \emph{bottom} panel the
 effective solid angle multiplied $d^{-2}$ and $d^{-4}$.}
\label{fig:omega}
\end{figure}

The efficiency of our search is summarized in the top panel of Figure~\ref{fig:omega}, as a function of KBO size. We also plot the
corresponding effective solid angle $\aeff(d)$, defined as:
\begin{equation}
\aeff(d) = \sum_*
\frac{H(d,\theta_*)}{D}~\frac{v_\mathrm{rel}}{D}~T_*~\epsilon(d, ~\theta_\star),
\end{equation} where $H(d, \theta_*)$ is the cross section of the event,
 which depends on both the diameter of the KBO and the star angular
 size as indicated by $\theta_*$; $v_\mathrm{rel}$ is the relative velocity of
 the KBO, which depends on the elongation angle which is close to
 opposition for all of our observations; $D$ is the distance to the
 occulter (assumed to be $D=40~\mathrm{AU}$), $T_*$ the exposure for
 the star target (duration of the lightcurve), and $\epsilon(d,~\theta_\star)$ the
 recovery efficiency for that diameter: $\epsilon(d,~\theta_\star)~=~1$ if the implanted event was recovered, 0 otherwise. The sum is carried out over
 all of our lightcurves with $SNR \geq 25$. $\aeff(d)$ represents the
 equivalent sky coverage of our survey for targets at diameter $d$,
 accounting for a partial efficiency. The center panel of
 Figure~\ref{fig:omega} shows the effective solid angle as a function
 of diameter. The bottom panel shows the effective solid angle
 multiplied by bracketing slopes for the size distribution: $d^{-4}$ and
 $d^{-2}$, and it indicates the survey expects to see the
 largest number of detections near $d ~=~700~\mathrm{m}$.

\subsection{Rejection of false positives}
\begin{figure}[bt]
\centerline{\includegraphics[width=0.4\textwidth]{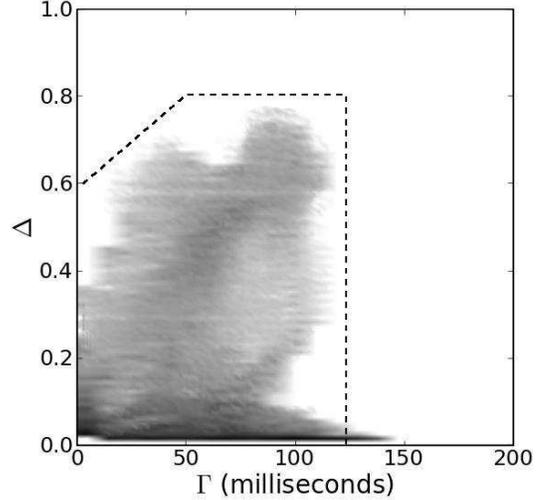}}
\caption[]{Phase space plot showing the regions of the \emph{flux
   decrease--duration} space occupied by occultations by KBOs of
 diameter 0.1 to 3~km, as observed through the MMT/Megacam system
 bandpass. We simulated occultations from KBOs in the size regime
 0.1--3.0~km, and we fit the occultations with an inverted top-hat
 function with parameters $\Delta $ and $\Gamma$. The intensity of the
 gray scale reflects the number of simulated occultations with best
 fit value $\Delta$ and $\Gamma$: white areas are void of occultations. }
\label{fig:occreg}
\end{figure}

At this stage we have more than a thousand candidates.  However, most
of the false positives can be removed in an automated fashion: we
reject fluctuations that appear simultaneously in more than one
lightcurve; those are most likely due to local weather or atmospheric
patterns that were not corrected in the de-trending phase because they
only affected a subset of lightcurves, and can be ruled out as
serendipitous occultations. We also reject any fluctuation that does
not have the right combination of depth and width. We empirically
investigate the relationship between the depth and the width of an
occultation by a KBO, as it is seen by our system, taking advantage of
our simulations. To define the depth and width of the events we fit
synthetic occultation lightcurves with inverted top-hat functions with
parameters $\Delta$ (depth) and $\Gamma$
(width). Figure~\ref{fig:occreg} shows the best fit values $\Delta$
and the $\Gamma$ for occultations simulated in the diameter
range $d=0.1~\mathrm{km}$ to $d=3.0~\mathrm{km}$, impact parameters
$b=0$ to $H/2$ and magnitude range 8 to 11 for F0V stars (the same set
that we used for our implantation with additional occultations from
objects $d<0.5~\mkm$).  The shaded region represents the area of this
phase space where at least one occultation was best fit by parameter
values $\Delta$ and $\Gamma$ (and the intensity of the shade reflects
the frequency of $\Delta-\Gamma$ best fits).  We can automatically
reject events outside the dashed polygon as incompatible with
$d\leq3.0~\mkm$ KBO occultations.\footnote{Note that the duration
  regime over which we recover events extend as far out as our largest
  window: $W=61$~points or 300~ms.} We are not sensitive to events
shallower than a 10\% flux drop.

 At this point the absolute number of candidates is small ($25$).  The
 remaining candidates are inspected visually (using DS9,
 \citealt{2003ASPC..295..489J}), and the lightcurves are extracted with a
 different photometric method (based on IRAF). All remaining
 candidates prove to be artifacts, mostly due to photometry.  No
 candidates are left after this elimination process.
\begin{figure*}[!t]
\centerline{\includegraphics[width=0.5\textwidth]{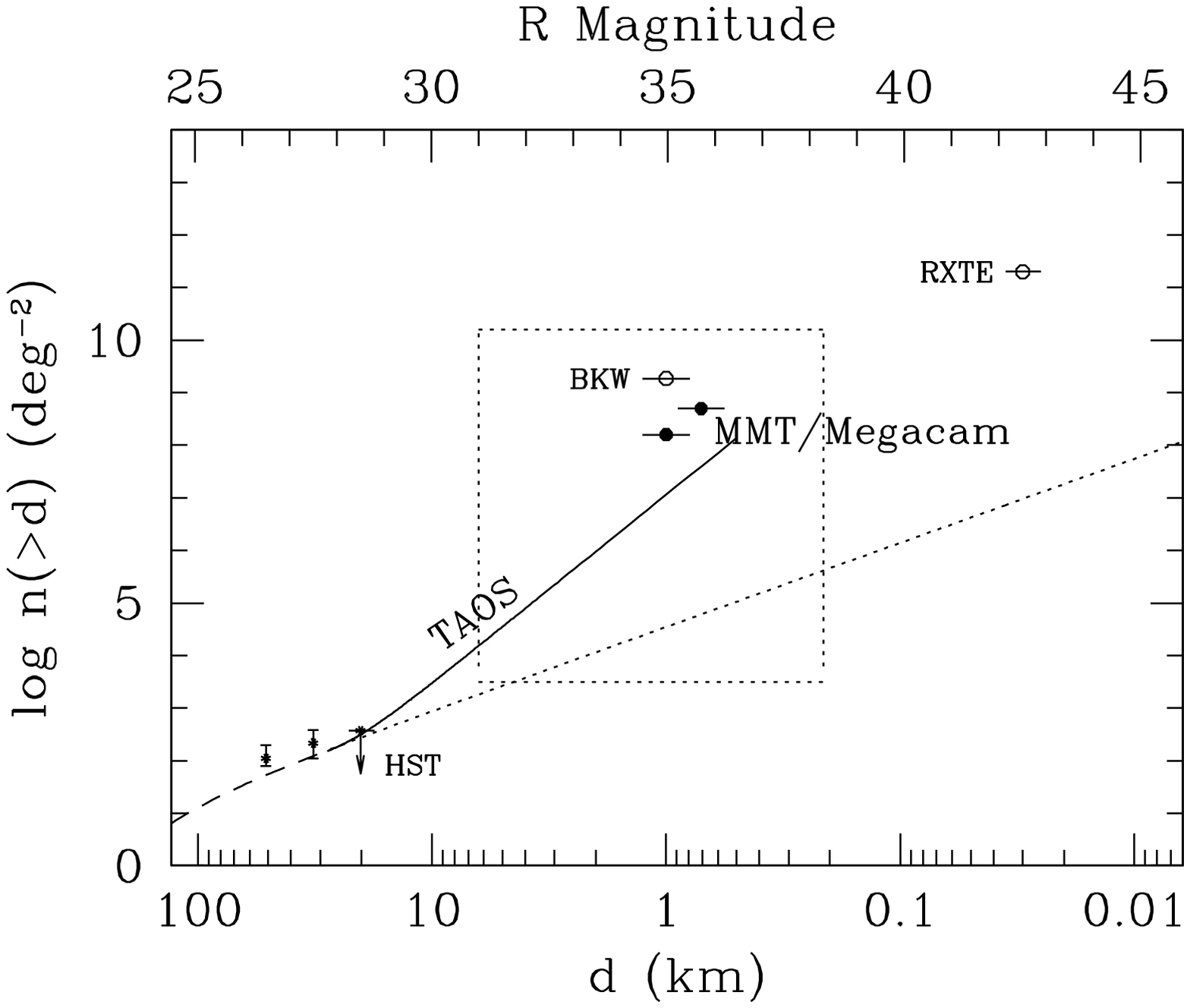}
\includegraphics[width=0.5\textwidth]{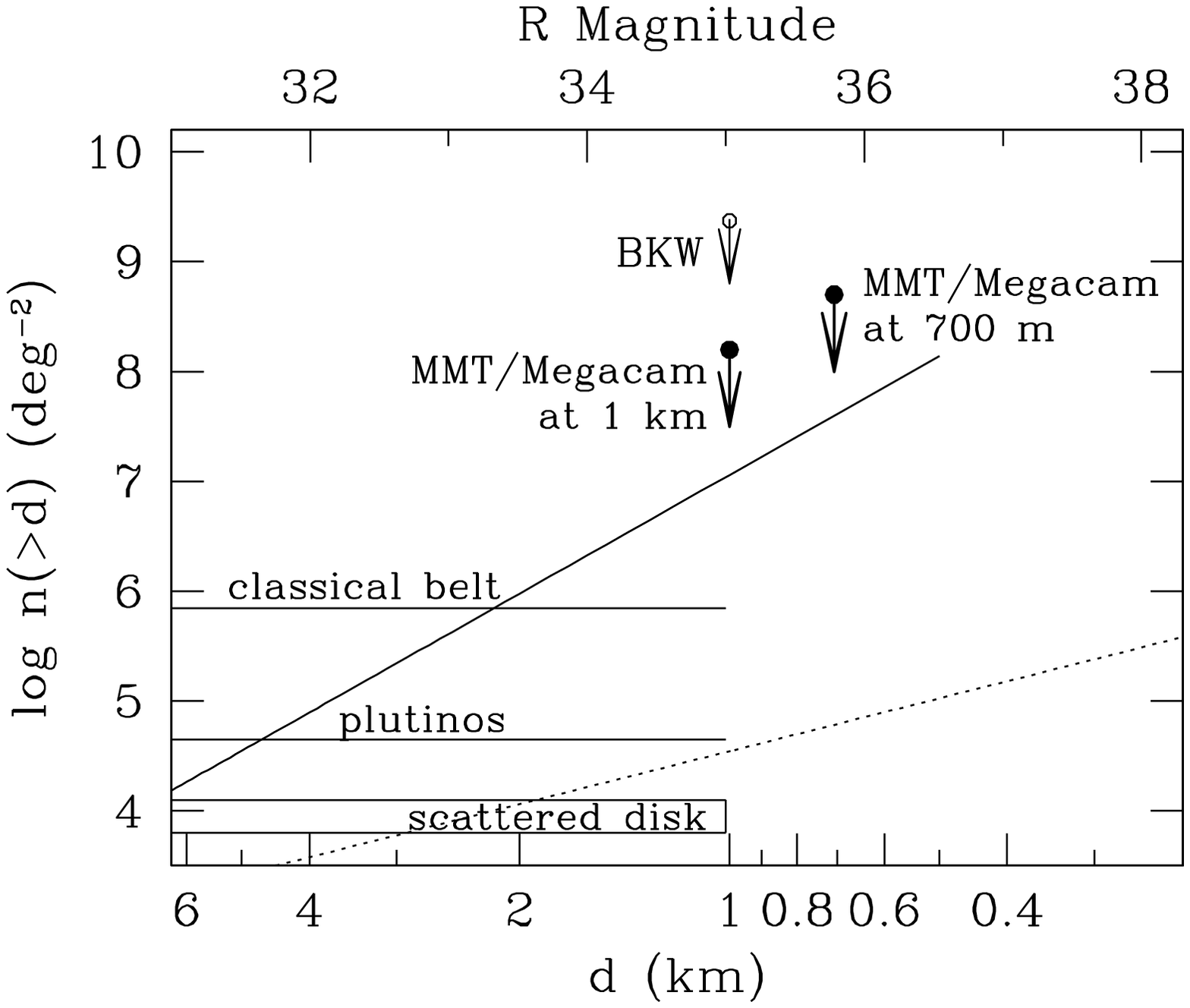}}
\caption[]{Upper limits to the surface density of KBOs. \emph{Left
   panel.} The dashed line is the best fit to the
 \citet{2004AJ....128.1364B} survey, extrapolated to
 $d=0.01\mkm$. Three data-points reported by
 \citet{2004AJ....128.1364B} are plotted (HST, the faintest data from
 direct observations). The straight line is the TAOS upper limit to
 the slope of the small size end size distribution: $q~<~4.6$
 \citep{2008ApJ...685L.157Z}. The result by
 \citet{2008AJ....135.1039B} is shown as an empty circle (BKW) as
 well as the X-ray result \citep[RXTE]{2008ApJ...677.1241J}.  The
 upper limits set by our survey at $d~=~1~\mathrm{km}$ and
 $d~=~0.7~\mathrm{km}$ are shown as filled circles.  The region
 relevant to our limit, enclosed in the square, is magnified on the
 right hand panel. \emph{Right panel.} Upper limits to the surface
 density of KBOs, zooming in the 0.2--6~km region of the size
 spectrum where our survey can place limits. Symbols and labels are
 the same as for the left panel.  The lower limits for JFC precursor
 populations are also shown \citep{1997Icar..127...13L,
   1997Icar..127....1M, 2008ApJ...687..714V}.}
\label{fig:uplim}
\end{figure*}

\section{Upper limit to the size distribution of KBOs and scientific interpretation}\label{sec:limits}
We now compare our limit to the size distribution of KBOs to that of \citet{2008AJ....135.1039B}.
\citet{2008AJ....135.1039B} derived an upper limit to the surface density of KBOs of diameter $d\ge1~\mathrm{km}$. They considered the data obtained by their own survey together with the data published by \citet{2006AJ....132..819R} and \citet{2007MNRAS.378.1287C}, assuming 100\% efficiency for each survey at $1~\mathrm{km}$, and obtaining a total effective coverage $\aeff = 5.4 \times 10^{-10} ~\mathrm{deg}^{2}$. The cross section $H$ used to calculate $\aeff$ is set to validate the 100\% efficiency assumption on a survey by survey basis.
 Our survey adds $7.0\times 10^{-9} ~\mathrm{deg}^2$ to the collective $\aeff$, allowing us to derive a limit over an order of magnitude stronger than the limit set by \citet{2008AJ....135.1039B}. Thus we set a comprehensive 95\% confidence level upper limit on the surface density of $d\ge1~\mathrm{km}$ KBOs at 40 AUs of $\Sigma_N (d\ge 1~\mathrm{km}) ~\sim~ 2.0 \times 10^8 ~\mathrm{deg}^{-2}$.

We can also derive a new upper limit for objects as small as
$700~\mathrm{m}$, where our efficiency is $\epsilon~(~d~=~700~\mathrm{m})~\sim~10\%$.  We
can set a 95\% confidence upper limit of $\Sigma_N (d\ge
0.7~\mathrm{km}) ~\sim~ 4.8 \times 10^8 ~\mathrm{deg}^{-2}$. These
limits are shown in Figure \ref{fig:uplim}, along with the TAOS
model-dependent upper limit and the limit set by the RXTE X-ray
survey.

\subsection{Comparison with the results from the TAOS survey}
Our survey aspires to be complementary to TAOS in that it potentially
could detect objects as small as 300~m. However, at this stage of our
work we are unable to push the detection limit below the
TAOS sensitivity ($500~\mathrm{m}$). Note that the recovery efficiency
for TAOS at 700 m is $\epsilon_\mathrm{TAOS} \sim 0.3\%$, a factor of four
lower than our efficiency.

The TAOS upper limit to the surface density of KBOs is presented as a
model-dependent limit, under the assumption of a straight power-low
behavior for the small end of the Kuiper Belt size distribution; it
is therefore not trivial to relate the two results, but it is clear
that the number of star-hours a dedicated survey can collect
compensates for the loss in efficiency at the small size end, and TAOS
is able to produce more stringent limits than our own. Our survey
would however capture the details of the diffraction feature with
exquisite sampling, while the information contained in the same
occultation, as observed by TAOS, would be greatly reduced due to the
slower sampling.  This would allow us to set constraints on the size
and distance of the occulter, while the size-distance-impact parameter
space is highly degenerate in the TAOS data.

\subsection{The Kuiper Belt as reservoir of Jupiter Family Comets}\label{sec:JFC}
The classical
Kuiper Belt, the scattered disk objects and the plutinos have all been
considered in dynamical simulations as possible reservoir of JFCs (see
\citealt{2008ApJ...687..714V} and references therein). The inclination
distribution of the JFCs strongly suggests a disk-like progenitor
population, favoring the Kuiper Belt over the Oort Cloud. Giant
planets generate long term gravitational perturbations that causes
weak orbital chaos, which explains the injection of comets to the JFCs
region \citep{1993AJ....105.1987H, 1995AJ....110.3073D,
1997Icar..127...13L}. The efficiency of this process depends on the
dynamical characteristics of the progenitor family.

Simulations of the injection process lead to lower limits on the
number of progenitors, which we can compare with our upper limit to
the surface density of KBOs.
\citet{2004AJ....128.1364B} discussed constraints on the progenitors of
the JFCs on the basis of their HST/ACS survey. This survey is however
only sensitive to objects greater than $\sim20~\mathrm{km}$ in
diameter, while the precursors of the JFCs are likely to be in the
size range $1-10~\mathrm{km}$. This is the typical observed size
of JFCs \citep{2008ssbn.book..397L} and it is likely that its
progenitor population would consist of objects of similar size (or
slightly larger) than the JFCs themselves.

In Figure \ref{fig:uplim} we show the lower limits to the KBO
populations (classical belt and plutinos) and scattered disk derived from
dynamical simulations. We use the estimate of
\citet{1997Icar..127...13L} for a population of cometary precursors
entirely in the classical Kuiper Belt, of \citet{1997Icar..127....1M}
for plutinos progenitors, and of \cite{2008ApJ...687..714V} for a
progenitor population in the scattered disk. As in
\citet{2004AJ....128.1364B} we convert the population estimates for
the Kuiper Belt populations into a surface density by assuming for each
population a projected sky area of $10^4~\mathrm{deg}^2$. \cite{2008ApJ...687..714V} provide information on the
fraction of time the objects in their simulation spend between 30 and
50 AUs and within $3\degree$ of the ecliptic plane, and these
fractions are used to calculate the minimum surface density of scattered disk
objects expected in the region of sky typically observed by occultation surveys.

We are not presently
able to exclude any of these populations as
progenitors of the JFCs. Future occultation surveys, with improved sensitivity, should
provide valuable information on the origin of JFCs.

\section{Conclusions and Future Work}\label{sec:conc}
We have devised a new observational method which allows fast photometry
with large telescopes with standard CCD cameras. We are able to
achieve high photometric rates (200 Hz) on tens of targets
simultaneously. The data reduction techniques for this kind of data
are still under development. The amplitude of our noise is
typically larger than the Poisson noise, and it
displays obvious deviations from normality. However, we prove this
method is suitable for gathering a large amount of precision fast
photometric data in few observing hours. We present a result that
lowers the upper limit set by similar sub-km target occultation surveys
by more than one order of magnitude for KBOs $d\ge1~\mathrm{km}$, and we can push the upper limit to $d\ge700~\mathrm{m}$. We confirm the result obtained by
dedicated Kuiper Belt occultation surveys.

The high speed sampling achieved with continuous readout mode will
enable the resolution of the diffraction features of any candidate
events, which is not possible with the TAOS project due to the lower
sampling rate they use.  This will allow us to set tight
constraints on the physical characteristics of an occulting system,
possibly breaking the degeneracy between impact parameter, size and
distance for sub-km KBOs. Furthermore, continuous readout mode enables
the simultaneous monitoring of as many as 100 stars, which is a
distinct advantage over the surveys of \citet{2006AJ....132..819R} and
\citet{2008AJ....135.1039B}, where only two stars can be sampled at a
time.

This observational technique has proven useful in testing telescope
performance and addressing guiding issues and it was used at the MMT to
test the drive servos.  Furthermore this observational method is a
promising technique for ground-based high precision photometry of
bright sources with large telescopes as it addresses many issues
typically encountered in observing bright targets \citep{2008arXiv0806.4911G}. Saturation is avoided without resorting to defocussing, it
involves no overhead due to readout and with a camera like Megacam,
with a large field of view, it allows the observation of many stars at
a time, guaranteeing the presence of a good number of comparison stars
that can be used to achieve high precision relative
photometry.

Further improvements in SNR might be achieved: we are exploring a
fitting photometry package that uses the Expectation--Minimization
algorithm, treating each row as a mixture of Gaussians, to better
separate the contribution from different sources. A possible way to
address the contamination due to unresolved sources is to subtract the
contribution from known unresolved sources (identified from the
stare--mode image, see Section \ref{sec:extraction}) using the trends
identified in the detrending phase (Section
~\ref{sec:detrending}). Another possibility is to de-trend the
lightcurves recursively, while allowing a variable phase offset.
Finally, it shall be noticed that Megacam will become available for
observations at the Magellan Clay Telescope, from where our target
fields, at the intersection of the galactic and ecliptic plane, could
be observed at a higher elevation. This would help reduce the noise
introduced by cross contamination and differential image motion, as the
athmospheric effects we encounter observing at high air masses would
be reduced.

\acknowledgments This work was supported in part by the NSF under
grant AST-0501681 and by NASA under grant NNG04G113G.  Observations
reported here were obtained at the MMT Observatory, a joint facility
of the Smithsonian Institution and the University of Arizona. This
research has made use of SAOImage DS9, developed by Smithsonian
Astrophysical Observatory.


\end{document}